\def\apj{{\it Astorphys. J.}}%
\def\ao{{\it Appl.~Opt.}}%
\def\aap{{\it Astron. Astrophys.}}%
\def\icarus{{\it Icarus}}%
\def\sun{\odot}%
\def\star{\ast}%

\documentclass[final]{eps}

\renewcommand \sun{\odot}
\newcommand\micron{\mbox{$\mu$m}}

\usepackage{graphicx}
\usepackage{natbib}
\usepackage{amsmath,amssymb}
\usepackage{times}

\volumenumber{xx}
\publishyear{2011}
\frompage{\pageref{firstpage}}
\topage{\pageref{finalpage}}

\shorttitle{H. Kobayashi \lowercase{\textit{et al.}}: Sublimation Temperature and Ring Formation of Circumstellar Dust Particles.}

\title{Sublimation Temperature of Circumstellar Dust Particles and Its Importance for Dust Ring Formation}
\author{Hiroshi Kobayashi$^1$, Hiroshi Kimura$^2$, Sei-ichiro Watanabe$^3$, Tetsuo Yamamoto$^4$, and Sebastian M\"uller$^1$}
\affiliation{$^1$Astrophysical Institute and University Observatory,
Friedrich Schiller University Jena, Schillergaesschen 2-3, 07745 Jena,
Germany\\
             $^2$Center for Planetary Science, c/o Graduate School of
	     Science of Kobe University, Nada-ku Rokkodai-cho 1-1, Kobe
	     657-8501, Japan\\
             $^3$Department of Earth and Planetary Sciences,
Graduate School of Environmental Studies, Nagoya University 
Furo-cho, Chikusa-ku, Nagoya, 464-8601, Japan\\
             $^4$Institute of Low Temperature Science, Hokkaido University 
Kita-Ku Kita 19 Nishi 8, Sapporo 060-0819, Japan}
\received{xxxx xx, 2010}\revised{xxxx xx, 2011}\accepted{xxxx xx, 2011}
\abstract{Dust particles in orbit around a star drift toward the
central star by the Poynting-Robertson effect and pile up by
sublimation. We analytically derive the pile-up magnitude, adopting a
simple model for optical cross sections. As a result, we find that the
sublimation temperature of drifting dust particles plays the most
important role in the pile-up rather than their optical property does.
Dust particles with high sublimation temperature form a
significant dust ring, which could be found in the vicinity of the sun
through in-situ spacecraft measurements.  While the existence of such a
ring in a debris disk could not be identified in the spectral energy
distribution (SED), the size of a dust-free zone shapes the SED. 
Since
we analytically obtain the location and temperature of sublimation,
these analytical formulae are useful to find such sublimation
evidences.}
\keywords{Sublimation - Dust - Interplanetary medium - Debris disks - 
Celestial mechanics.}

\begin{document}
\label{firstpage}
\maketitle
\copyrighttext{}

\section{Introduction}

Refractory dust grains in orbit around a star spiral into the star by
the Poynting-Robertson drag (hereafter P-R drag) and sublime in the
immediate vicinity of the star.  Because the particles lose their mass during
sublimation, the ratio $\beta$ of radiation pressure to gravity of the
star acting on each particle ordinarily increases.  
As a result, their radial-drift rates decrease 
and the particles pile up at the outer edge of their sublimation zone
\citep[e.g.,][]{mukai79,burns79}.  This is a mechanism to form a dust
ring proposed by \citet{belton66} as an accumulation of interplanetary dust
grains at their sublimation zone.  Ring formation of drifting dust
particles is not limited to refractory grains around the sun but it also
takes place for icy grains from the Edgeworth-Kuiper belt 
and for dust in debris disks \citep{kobayashi08,kobayashi10}. 
Therefore, dust ring formation due to sublimation of dust particles is a
common process for radially drifting particles by the P-R drag. 

The orbital eccentricity and semimajor axis of a dust particle evolve by
sublimation due to an increase in its $\beta$ ratio 
as well as by
the P-R drag.  
We have derived the secular evolution
rates of the orbital elements \citep{kobayashi09}. 
The derived rates allow us to find an analytical solution of the
enhancement factors for the number density and optical depth of dust
particles due to a pile-up caused by sublimation. Our analytical
solution is found to reproduce numerical simulations of the pile-up well 
but its applicability is restricted for low eccentricities of subliming
dust particles.  
The analytical solution shows that the enhancement factors depend on
dust shapes and materials as expected from previous numerical studies
\citep[cf.][]{kimura97}. 
Although the solution includes physical quantities for
the shapes and materials, it does not explicitly show which quantity essentially
determines the enhancement factors.

The goal of this paper is to derive simplified formulae that explicitly
indicate the dependence of dust ring formation on materials and
structures of dust particles.  In this paper, we adopt a simple model
for the optical cross sections of fractal dust particles and
analytically obtain not only the enhancement factors but also the
location of the pile-up and sublimation temperature.  In addition, we
extend the model of \citet{kobayashi09} 
by taking into account
orbital eccentricities of subliming dust particles.

In Section 2,
we derive the sublimation temperature as a function of the latent heat.  
In Section 3, we introduce the characteristic radius of fractal dust and 
derive the sublimation distance for that dust. 
In Section 4, we simplify the formulae of enhancement factors derived by
\citet{kobayashi09} 
and obtain the new formulae that show
explicitly the dependence on materials and structures of the particles. 
We provide a recipe to use our analytical formulae in Section 5, apply our simplified formulae to both the
solar system and extrasolar debris disks, and discuss observational
possibilities of dust sublimation in Section 6. We summarize our
findings in Section 7.

\section{Sublimation temperature}
\label{sc:temp}

We consider dust particles in orbit around a central star with mass
$M_\star$. Driven by the P-R drag, they drift inward until 
they actively sublime in the vicinity of the star. 
We have shown in \citet{kobayashi09} 
that the ring formation due to
sublimation occurs only for their low orbital eccentricities $e$ and obtained the secular change of semimajor axis $a$
of the particle with mass $m$ as 
\begin{eqnarray}
 \left< \frac{da}{dt} \right>
 &=& - \eta 
 \frac{\beta}{1-\beta}
 \frac{a}{m} \left. \frac{dm}{dt}\right|_{r=a} 
 - \frac{\beta G M_\star}{c} \frac{2}{a},
\label{eq:da_low}
\end{eqnarray}
where $\eta \equiv - \ln \beta / \ln m$, $-dm/dt|_{r = a}$ is
the mass-loss rate of the particle at the distance $r = a$, $G$ is
the gravitational constant, and $c$ is the speed of light. 
The $\beta$ ratio is given by 
\begin{equation}
 \beta = \frac{L_\star {\bar{C}}_{\rm pr}}{4\pi cGM_\star m}, 
\label{eq:beta}
\end{equation}
where 
$\bar C_{\rm pr}$ is the radiation
pressure cross section averaged over the stellar radiation spectrum and
$L_\star$ is the stellar luminosity. The
first and second terms on the right-hand side of Eq.~(\ref{eq:da_low})
represent 
the drift rates due to sublimation and the P-R drag, respectively.
Although we consider only the P-R drag from stellar radiation, 
the P-R drag due to stellar wind also transports the particles. 
However, the magnitude
of pile-up, its location, and sublimation temperature 
hardly depend on which drag determines their transport  
\citep{kobayashi08,kobayashi09}. 

A particle generated in a dust source initially spirals toward a
star by the P-R effect.  As it approaches the star due to the P-R
inward drift, its temperature rises high and it finally 
starts active sublimation.  The drift turns outward by sublimation when
$\beta$ increases with mass loss.  The radial motion of the particles
becomes much slower than the P-R drift alone, resulting in a pile-up
of the particles.  Note that other mass-loss mechanisms such as
sputtering by stellar wind and UV radiation are negligible during active
sublimation.\footnote{\footnotemark[1]We consider dust particles that can drift into their active sublimation
zone.  This is valid in the solar system, since the size decreasing
timescale due to sputtering is longer than the drift time due to the
P-R effect \citep{mukai81}.  However, we note that icy particles may not come to
their sublimation zone around highly luminous stars because of strong UV
sputtering \citep{grigorieva}. 
}

The mass loss rate of a particle due to sublimation is given by
\begin{equation}
 -\frac{dm}{dt} = A \sqrt{\frac{\mu m_{\rm u}}{2 \pi kT}} 
   P_0(T) \exp \left( - \frac{\mu m_{\rm u} H}{k T} \right),
\label{eq:ds}
\end{equation}
where $A$ is the surface area of the particle, $H$ is the latent heat of sublimation, $\mu$ is the mean molecular weight of the dust material, 
$m_{\rm u}$ is the atomic mass unit, and $k$ is the Boltzmann constant.  
Here the saturated vapor pressure at temperature $T$ is
expressed by $P_0(T) \exp ( - \mu m_{\rm u} H /k T ) $ with $P_0(T)$
being only weakly dependent on $T$.

During active sublimation, the first term on the right-hand side of
Eq.~(\ref{eq:da_low})
increases and then $\langle \dot a \rangle$ nearly vanishes.  The temperature $T_{\rm sub}$ at
active sublimation is approximately determined by $\langle \dot a
\rangle = 0$.  Substituting Eq.~(\ref{eq:ds}) into Eq.~(\ref{eq:da_low})
for $\langle \dot a \rangle = 0$, we have
\begin{equation}
 T_{\rm sub} = - \frac{\mu m_{\rm u} H}{k} 
\left[  \ln 
\left(\frac{2 GM_\star }{c a^2} 
 \frac{m}{A P_0} \frac{1-\beta}{\eta}
\sqrt{\frac{2\pi k T_{\rm sub}}{\mu m_{\rm
 u}}} 
\right)\right]^{-1}.\label{eq:temp_ori} 
\end{equation}
Although Eq.~(\ref{eq:temp_ori}) is a function of $a$ as well as $T_{\rm sub}$, 
the natural logarithmic function on the right-hand side has little sensitivity
to $a$ and $T_{\rm sub}$. 
Therefore, the sublimation temperature may be approximated by
\begin{equation}
 T_{\rm sub} \simeq 1.3 \times 10^3 \, \xi^{-1} \left(\frac{H}{3.2 \times 10^{10} \, {\rm
				erg\,g}^{-1}}\right)
 \Biggl( \frac{\mu}{170}\Biggr) \, {\rm K},\label{eq:temp_app} 
\end{equation}
where
\begin{equation}
 \xi = 1 + 0.02 \ln \left(\frac{P_0}{6.7\times 10^{14} \, 
			    {\rm dyn \, cm}^{-2}}\right). 
\end{equation}
Here, we set $m = 1.1\times 10^{-12}$\,g, $\beta = 1/2$, $\eta = 1/3$,  
$T_{\rm sub}=1300$\,K and $a = 15 R_\sun$ with the solar radius $R_\sun$ 
in the argument of the 
logarithmic function under the assumption of a spherical olivine 
dust particle around the sun; the other
choice of these values does not change the result significantly because of the
slowly-varying properties of the logarithmic function.  

\begin{figure}[t]
 \centerline{\includegraphics[width=7cm]{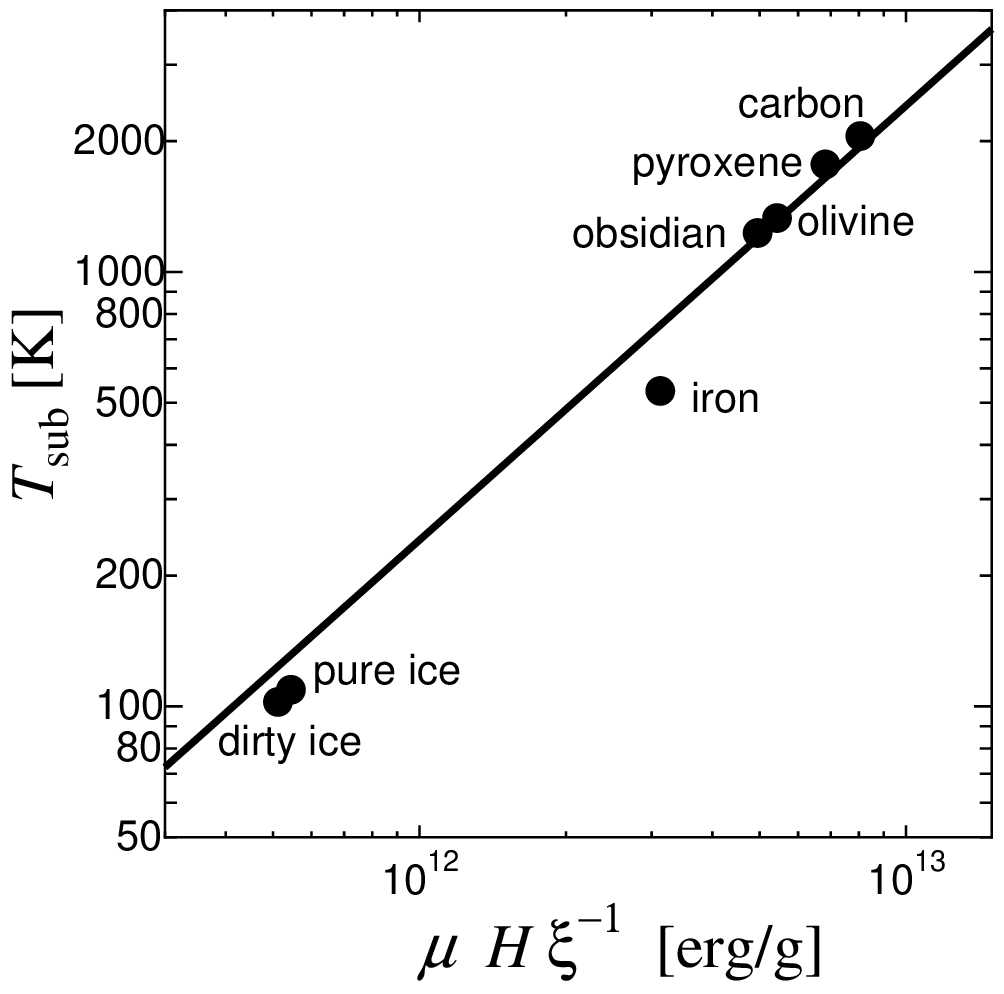}} \caption{
Sublimation
temperature $T_{\rm sub}$ as a function of $\mu H \xi^{-1}$, where $\xi
 = 1 +0.02 \ln (P_0/6.7 \times 10^{14} \,{\rm dyn \, cm}^{-2})$. 
The mean molecular weight $\mu$, the latent heat
 $H$, and the vapor pressure in the limit of high temperature, $P_0$, are listed
in Table 1.  The solid line indicates Eq.~(\ref{eq:temp_app}). Circles
represent $T_{\rm sub}$ obtained from the method of
\citet{kobayashi09} around the sun.  
\label{fig:ts}}
\end{figure}

Equation (\ref{eq:temp_app}) indicates that the active sublimation temperature
is mainly determined by the latent heat of sublimation and mean molecular
weight of the particles.  
This explains the findings by \citet{kobayashi08} that the temperature is insensitive to the stellar parameters, $M_\star$ and $L_\star$. 
As a consistency check, 
we calculate the sublimation temperatures according to 
\citet{kobayashi09} 
for materials listed in Table 1 
and compare the temperatures with
Eq.~(\ref{eq:temp_app}) (see Fig.~\ref{fig:ts}).  In spite of the
simplification, Eq.~(\ref{eq:temp_app}) is in good agreement with the
temperature given by the procedure of 
\citet{kobayashi09}. 

\begin{table*}[t]
\renewcommand{\arraystretch}{1.2}
\vspace{-.3cm}
 \caption{Material parameters: the material density, $\mu$ is
 the mean molecular weight, $H$ is the latent heat of sublimation, and $P_0$
 is the saturated vapor pressure $P_{\rm v}$ in the limit of high
 temperature.}
\vspace{-.1cm}
  \begin{center}
   \begin{tabular}{c c c c c }
    \\
    \hline
    & material density [g ${\rm cm}^{-3}$]& $\mu$& $H$ [erg ${\rm g}^{-1}$] & $P_0$
    [dyn ${\rm cm}^{-2}$] \\
    \hline
    olivine & 3.3   & 169.1 & $3.21 \times 10^{10}$  & $6.72 \times
		    10^{14}$\\
    pyroxene &3.3& 60.1 & $9.60 \times 10^{10}$ &$3.12\times 10^{11}$ \\

    obsidian  & 2.37   & 67.0 & $7.12 \times 10^{10}$ & $1.07 \times 10^{14}$\\
		    
    carbon & 1.95   & 12.0 & $7.27 \times 10^{11}$ & $4.31 \times 10^{16}$\\
		    
    iron & 7.86   & 55.8 & $2.97\times 10^{10}$$^{\,\rm a}$  & $5.00\times
		    10^4$$^{\,\rm a}$\\
    pure ice & 1.0  & 18 & $2.83 \times 10^{10}$$^{\,\rm
		b}$  & $3.08 \times 10^{13}$ $^{\rm b}$\\
    dirty ice & 1.4   & 18 & $2.83 \times 10^{10}$$^{\,\rm
		b}$  & $2.67 \times 10^{13}$ $^{\rm b}$\\
    \hline
    \\
   \end{tabular}\\ 
  \end{center}
Note --- obsidian is formed as an igneous rock and may not be plausible
 as interplanetary dust, but is applied for a comparison
 with previous studies \citep[e.g.,][]{mukai79}. 
\\
$^{a,b}$ $H$ and $P_0$ are obtained from the following formulae with the
 sublimation temperature of each material. \\
 $^{\rm a}$ $P_{\rm v} = 1.33\times10^4\exp(-2108/T+16.89-2.14\ln T)$
 $\,{\rm dyn \, cm}^{-2}$ 
\citep{lamy}
\\
 $^{\rm b}$ $\log P_{\rm v} = -2445.5646/T + 8.2312 \log T - 0.01677006 T 
 + 1.20514 \times 10^{-5} T^2 - 3.63227 $  
\citep[$P_{\rm v}$ in the cgs unit;][]{washburn} 
\end{table*}

\section{Fractal dust approximation}
\label{sc:fractal_dust}

We introduce the characteristic radius $s$ of a dust particle, which is
defined as
\begin{equation}
 s^2 = \frac{5}{3}\frac{\int \rho_{\rm i} \tilde{s}^2 d V}{\int \rho_{\rm i}
  dV}, 
\end{equation}
where $\int dV$ means an integration over volume, $\tilde{s}$ is the 
distance from its center of mass, and $\rho_{\rm i}$ is its interior
density.  We consider that particles have a fractal structure; 
the mass-radius relation of the particles is given by $m \propto s^D$
for a constant fractal dimension $D$.  For the fractal dust,
$s= s_{\rm g} \sqrt{5/3}$, where $s_{\rm g}$ is the gyration radius of the dust
\citep{mukai92}. 
For homogeneous
spherical dust, the characteristic radius reduces to the radius of the
sphere.  The cross sections of scattering and absorption of light are
approximately described by a function of $\pi s^2$ and $2 \pi s/\lambda$, where $\lambda$
is the wavelength at the peak of light spectrum from the dust
\citep{mukai92}.

The smallest dust particles before active sublimation contribute most to the enhancements of
number density and optical depth at the pile-up \citep{kobayashi09}.
Small particles produced by parent bodies in circular orbits are expelled by
the radiation pressure if $\beta > 1/2$.  Thus, the minimum
characteristic radius $s_{\rm 0min}$ of dust particles prior to active
sublimation corresponds to $\beta = 1/2$.  The radiation pressure
cross section $\bar C_{\rm pr}$ is roughly given by $\pi s_{\rm 0min}^2$ in
Eq.~(\ref{eq:beta}) for $s_{\rm 0min} \gtrsim \lambda_\star$.  Then, we
have
\begin{equation}
 s_{\rm 0min} = 1.2 \, 
  \left(\frac{L_\star}{L_\sun} \right)\, \left(\frac{M_\sun}{M_\star} \right)
  \left(\frac{\rho}{1.0 \, {\rm g \,cm}^{-3}}\right)^{-1} \,\micron, 
  \label{eq:rc0min}
\end{equation}
where $M_\sun$ and $L_\sun$, respectively, denote the solar mass and
luminosity.  Note that $\rho = 3 m_{\rm 0min}/4\pi s_{\rm 0min}^3$ is
the effective density of a dust particle with the characteristic radius
$s_{\rm 0min}$ and mass $m_{\rm 0min}$ in the following derivation.  In
addition, we discuss the application limit of our formulae in Appendix
A.

\subsection{Sublimation distance 
\label{sc:dist}
}

We introduce the dimensionless parameter $x$, 
\begin{eqnarray}
x &=& \frac{2 \pi s_{\rm 0min}}{\lambda_{\rm sub}}
\nonumber\label{eq:x}
\\ 
 &=& 3.4 \, \left(\frac{L_\star}{L_\sun}\right)\,
\left(\frac{M_\sun}{M_\star}\right)\,
  \left(\frac{\rho}{1.0 {\rm g \,cm}^{-3}}\right)^{-1} \left(\frac{T_{\rm
      sub}}{1300\,{\rm K}}\right),  
\label{eq:x} 
\end{eqnarray}
where $\lambda_{\rm sub}$ is the wavelength at the peak of thermal
emission from subliming dust with temperature $T_{\rm sub}$.  We
approximate $\lambda_{\rm sub} = (2898 \, {\rm K} /T_{\rm sub})
\,\micron$, which is the wavelength at the peak of a blackbody radiation
spectrum with $T_{\rm sub}$.

Since we deal with dust dynamics in optically thin disks,
the equilibrium temperature $T$ of a dust particle at a certain distance
from a star is determined by energy balance among absorption of incident
stellar radiation and emission of thermal radiation. Therefore, the
relation between temperature $T$ and distance $r=a$ is approximately
given by \citep[e.g.,][]{kobayashi09} 
\begin{equation}
 \frac{L_\star{\bar{C}_\star}}{4 \pi a^2 } 
 = 4 {\bar{C}_{\rm d}} \sigma_{\rm SB} T^4, 
\label{eq:temp}
\end{equation}
if $a$ is much larger than the radius of the central star.  Here,
$\sigma_{\rm SB}$ is the Stephan-Boltzmann constant and $\bar
C_\star(s_{\rm 0min})$ and $\bar C_{\rm d}$ are the absorption
cross sections integrated over the stellar spectrum and the thermal
emission from the dust particle, respectively.

Because $s_{\rm 0min}$ is larger than $\lambda_\star$, the cross section
$\bar C_\star(s_{\rm 0min})$ is approximated by the geometrical cross
section;
\begin{equation}
\bar{C}_\star(s_{\rm 0min}) = \pi s_{\rm 0min}^2.\label{eq:C_s}  
\end{equation}
The cross section $\bar C_{\rm d}(s_{\rm 0min})$ may be $\pi s_{\rm
0min}^2$ for $x \gg 1$ and $\pi s_{\rm 0min}^2 x$ for $x \ll 1$.  We
connect them in a simple form as
\begin{equation}
 \bar{C}_{\rm d}(s_{\rm 0min}) = \pi s_{\rm 0min}^2 \left(\frac{ x}{1+x}\right).\label{eq:c_d} 
\end{equation}

When the temperature of the smallest drifting particles reaches $T_{\rm
sub}$,  their pile-up results in a peak on their radial distribution
\citep{kobayashi09}.  
With the application of the cross sections given by Eqs.~(\ref{eq:C_s}),
(\ref{eq:c_d}) to Eq.~(\ref{eq:temp}), the sublimation distance $a_{\rm
sub}$ at the peak is obtained as 
\begin{equation}
a_{\rm sub}  
 \simeq 
9.9\, \left( 1+x^{-1} \right)^{1/2}
\left(\frac{L_\star}{L_\sun} \right)^{1/2}
\left(\frac{T_{\rm sub}}{1300 \, {\rm K}}\right)^{-2} R_\sun, 
\label{eq:asub_app}
\end{equation}
where $R_\sun = 4.65 \times 10^{-3}$\,AU. 

\begin{figure}[t]
\centerline{\includegraphics[width=7cm,clip]{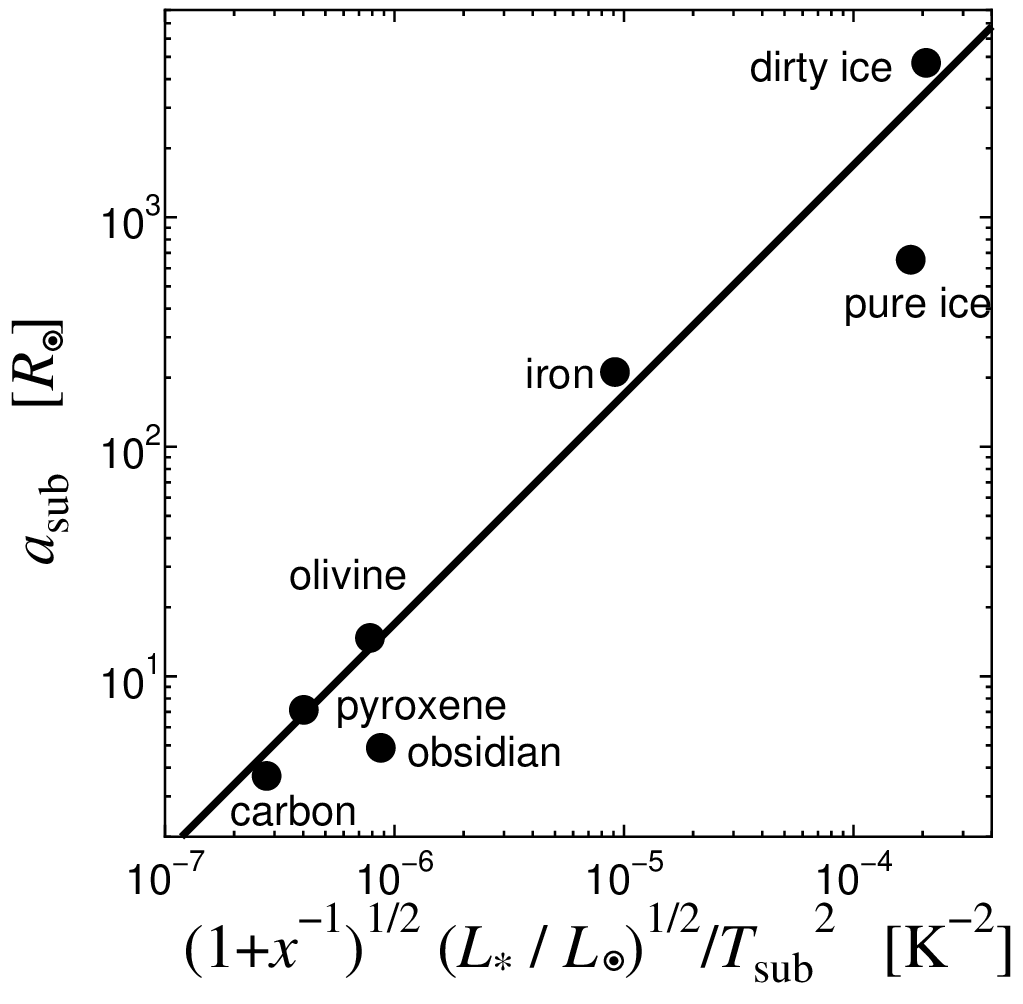}} 
\caption{Sublimation distance
$a_{\rm sub}$ in solar radii $R_\sun$ as a function of $(1+x^{-1})^{1/2}
(L_\star/L_\sun)^{1/2} /T_{\rm sub}^2$, where $x$ is given by
Eq.~(\ref{eq:x}).  The solid line indicates Eq.~(\ref{eq:asub_app}). 
Circles represent $a_{\rm sub}$ obtained from the method of
\citet{kobayashi09} around the sun.  \label{fig:asub}}
\end{figure}

Inserting $x$ given by Eq.~(\ref{eq:x}) in Eq.~(\ref{eq:asub_app}), we
have $a_{\rm sub} \propto L_\star^{1/2} T_{\rm sub}^{-2}$ for $x \gg 1$
and $a_{\rm sub} \propto M_\star^{1/2} T_{\rm sub}^{-3}$ for $x \ll 1$. 
In \citet{kobayashi08}, 
our simulations have shown this dependence for dirty ice under the
assumption that $L_\star \propto M_\star^{3.5}$.
We coupled
Eqs.~(\ref{eq:temp_ori}) and (\ref{eq:temp}) and adopted the
cross sections calculated with Mie theory\footnote[2]{\footnotemark[2]We apply the
complex refractive indices of olivine from \citet{huffman76} 
and \citet{mukai_koike} 
of pyroxene from \citet{huffman71}, 
\citet{hiroi}, 
\citet{roush91}, 
and \citet{henning}, 
of obsidian from \citet{lamy78} 
and \citet{pollack}, 
of carbon from 
\citet{hanner}, 
of iron form 
\citet{johnson} 
and \citet{ordal}, 
of pure and dirty ice form 
\citet{warren} 
and \citet{li}. 
}, and then obtained $a_{\rm
sub}$ \citep{kobayashi09}.  Equation~(\ref{eq:asub_app}) agrees well with $a_{\rm sub}$
derived from the method of \citet{kobayashi09} (see
Fig.~\ref{fig:asub}).  However, Eq.~(\ref{eq:asub_app}) overestimates
$a_{\rm sub}$ for less-absorbing materials (pure ice and obsidian)
because our assumption of $\bar C_{\star} = \pi s_{\rm 0min}^2$ is not
appropriate for such materials.  Nevertheless,
Eq.~(\ref{eq:asub_app}) is reasonably accurate for absorbing or compound
dust (dirty ice).

\section{Enhancement factor}

Dust particles with mass $m_0$ in the range from $m_{\rm 0min}$ to $m_{\rm
0max}$ are mainly controlled by the P-R drag in their source and
therefore spiral into the sublimation zone.  
As mentioned above, the smallest drifting dust with $m_{\rm 0min}$
corresponds to $\beta = 1/2$.  If the drifting timescale of dust
particles due to the P-R drag $t_{\rm PR}$ is much shorter than the
timescale of their mutual, destructive collisions $t_{\rm col}$, 
the particles can get out of the dust source region by the P-R drag. 
The ratio of $t_{\rm PR}$ to $t_{\rm col}$ increases with mass or size. 
Large dust particles with $t_{\rm PR} \gtrsim t_{\rm col}$ are 
collisionally ground down prior to their inward drifts.
Therefore, the largest dust $m_{\rm 0max}$ considered here roughly
satisfies the  condition $t_{\rm PR} \sim t_{\rm col}$ at the source region. 

In the steady state, the number density of drifting dust particles is
inversely proportional to the drift velocity
\citep[e.g.,][]{kobayashi09}.  Since the drift velocity of dust
particles due to the P-R drag is proportional to $\beta$, their mass
distribution is affected by the mass dependence $\eta = - d \ln \beta /
d \ln m$.  If the differential mass distribution of the dust source is
proportional to $m^{-b}$, that of drifting dust is modulated to $m^{-b +
\eta}$ \citep[e.g.,][]{moromartin}.
Provided that successive collisions mainly produce dust particles in the dust
source, we have $b = (11+3p)/(6+3p)$ for the steady state of collisional
evolution, where
$v^2/Q_{\rm D}^* \propto m^{-p}$ \citep{kt}. 
Here,
$Q_{\rm D}^*$ is the specific impact energy threshold for destructive
collisions and $v$ is the collisional velocity.  From the hydrodynamical
simulations and laboratory experiments, $Q_{\rm D}^* \propto
m^{-0.2}$ to $m^0$ for small dust particles 
\citep{holsapple, benz99}. 
Since $p = - 0.2$ to 0 for a constant $v$ with mass, $b$ is
estimated to be 1.8--1.9. This means that the smallest particles contribute
most to the number density before dust particles start to actively
sublime, while the largest particles dominate the optical depth prior
to active sublimation. 

When the temperatures of dust particles reach $T_{\rm sub}$, they start
to sublime actively. Their $\dot a$ do not vanish perfectly, but they
have very small $|\dot a|$ relative to the initial P-R drift velocity. The magnitude of a pile-up due to
sublimation is determined by the ratio of these drift rates
\citep{kobayashi09}.  Because the drift rate at the sublimation zone is
independent of the initial mass and the initial P-R drift rate decreases
with dust mass, the initially small dust piles up effectively.  As a
result, both 
the number density and the optical depth at the sublimation zone are
determined by the initially smallest dust.

The number density is a quantity that can be measured by in-situ
spacecraft instruments, while the optical depth is a key factor for
observations by telescopes. In \citet{kobayashi09}, we have provided enhancement factors for the
number density and the optical depth due to sublimation.  
Here, we apply the
simple model for optical cross sections in Eqs.(\ref{eq:C_s}),
(\ref{eq:c_d}) and the properties of the fractal dust given by Eq.~(\ref{eq:fractal_dust_model}).  Furthermore, we take into account an increase of
eccentricities from $e_1$ due to active sublimation.  The number-density
enhancement factor $f_{{N}}$ and the optical-depth enhancement
factor $f_{\tau}$ at the sublimation zone are then given by (see Appendix B
for the derivation)
\begin{eqnarray}
  f_{N} &\simeq&  \frac{b-1-\eta}{b-1} 
   g(x) h(e_1) + 1,\label{eq:f_n} \\
f_{\tau} &\simeq& \frac{2-b}{b-1} 
\left(\frac{m_{\rm 0min}}{m_{\rm 0max}}\right)^{2-b} 
 g(x) h(e_1) + 1,  \label{eq:f_tau}
\end{eqnarray}
where the functions $g(x)$ and $h(e_1)$ include the dependence on $x$
and $e_1$, respectively. 
They are given by 
\begin{eqnarray}
 g(x) &=& \frac{2 \alpha I (1+x)}{2 \alpha (1+x) + I }\label{eq:gfunc} \\
 h(e_1) &=& 1-\left[2- (2 I e_1 )^{-1/I} \right]^{(b-1)/\eta}
\label{eq:he1}
\end{eqnarray}
where $\alpha = - d \ln \beta / d \ln s = D-2$ and $I = \mu m_{\rm u}
H/4kT_{\rm sub} \simeq 13$.  Since we assume that the mass differential
number of the drifting particles is proportional to $m_0^{-b+\eta}$
before active sublimation, the dependence of $f_\tau$ on $m_{\rm
0min}/m_{\rm 0max}$ seen in Eq.~(\ref{eq:f_tau}) differs from that of 
\citet{kobayashi09}. 
This mass distribution is more realistic and
consistent with that of dust particles measured by spacecraft around the
earth \citep{grun}. 
Equation~(\ref{eq:he1}) for $h(e_1)$ is
applicable for $e_1$ ranging from $1/ 2^{I+1} I \simeq 7 \times 10^{-6}$
to $1/2 I \simeq 0.05$.  Dust particles do not pile up for $e_1 > 1/ 2
I$ and hence we give $h(e_1) = 0$ for $e_1 > 1/ 2 I$ 
\citep{kobayashi09}. 
In addition, $h(e_1) = 1$ for $e_1 < 1/ 2^{I+1} I$, while
drifting dust particles hardly reach such small eccentricities ($e_1 < 1/
2^{\kappa+1} I\sim 10^{-5}$) because their eccentricities naturally rise
as high as the ratio of the Keplarian velocity to the speed of light $
[\sim 10^{-4} (a/{\rm 1\,AU})^{-1/2} (M_\star/M_\sun)^{1/2}]$ by the P-R
effect.

\begin{figure}[t]
\centerline{\includegraphics[width=7cm]{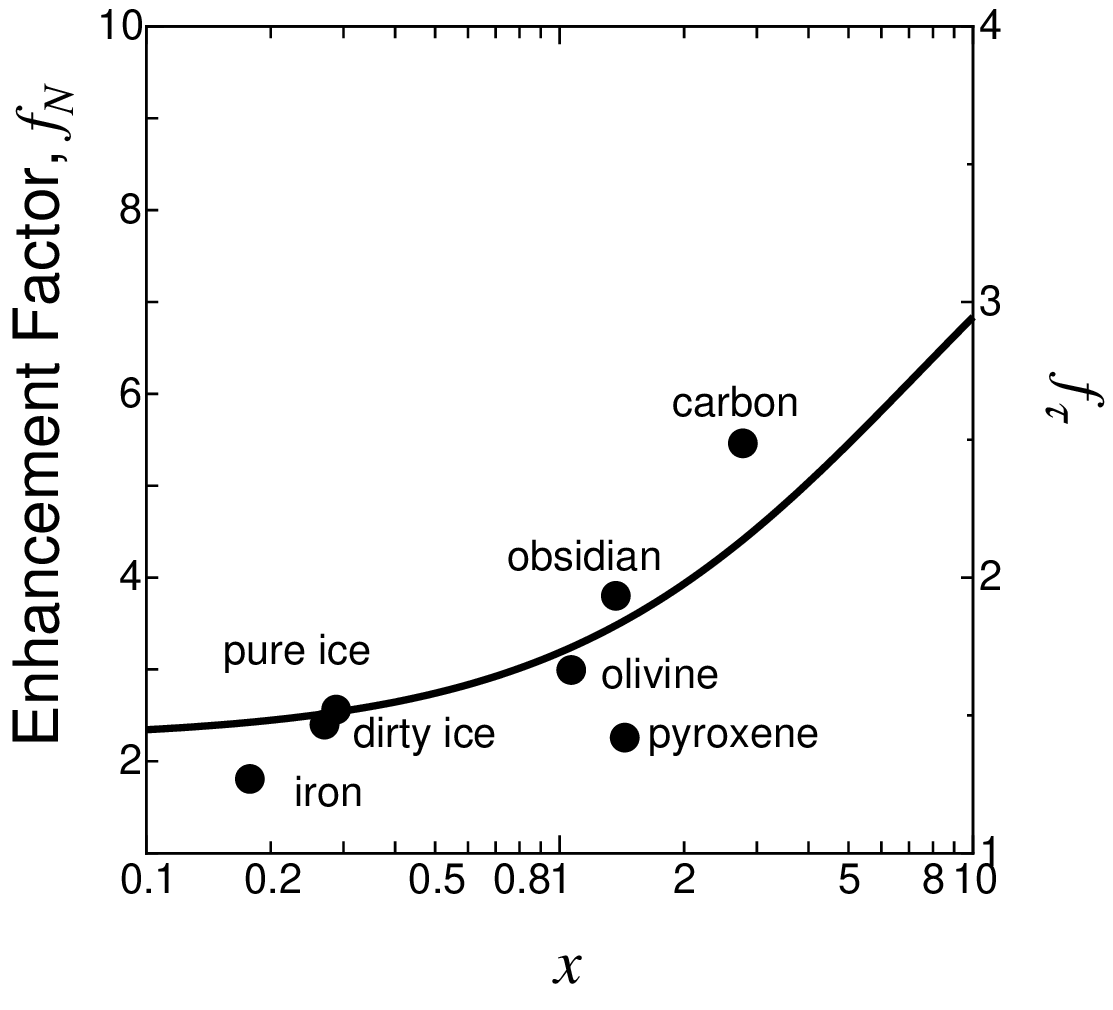}} \caption{ The enhancement
factors for low orbital eccentricities as a function of $x$, where the
dimensionless parameter $x$ is determined by Eqs.~(\ref{eq:x}).  Solid
line represents Eq.~(\ref{eq:f_n}) and (\ref{eq:f_tau}) with a use of
$m_{\rm 0min} = m_{\rm 0max}$ and $h(e_1) = 1$.  Circles indicate the
factor numerically calculated by Eq.~(68) and (69) of
\citet{kobayashi09} around the sun for spherical dust listed in Table 1.
\label{fig:enhan_temp} }
\end{figure}

In Fig.~\ref{fig:enhan_temp}, we compare the simplified formulae given by
Eqs.~(\ref{eq:f_n}), (\ref{eq:f_tau}) with the enhancement factors 
rigorously 
calculated by the formulae of \citet{kobayashi09}. 
The $x$ dependence of
the enhancement factors given by Eqs.~(\ref{eq:f_n}), (\ref{eq:f_tau})
is shown in the function $g(x)$, which is an increasing function ranging from
$2 \alpha $ ($x = 0$) to $I$ ($x = \infty$). Equations~(\ref{eq:f_n}), (\ref{eq:f_tau}) briefly
explain the tendency of the enhancement factors; the materials with high
$x$ produce high enhancement factors.  

Because $x$ is proportional to $T_{\rm sub}/\rho$ 
(see Eq.~(\ref{eq:x})), the enhancement factors increase with $T_{\rm
sub}/\rho$.  Thus, materials with high sublimation temperature tend to
pile up sufficiently. In addition, fluffy dust particles with $D \simeq 2$ cannot effectively
pile up even though $\rho$ is low \citep{kimura97}. 
This is
explained by the low $\alpha = D-2$ in the function $g(x)$. 
Particles with $D = 3$ produce the highest enhancement factors.  
In spite of $D=3$, compact particles are not the best for the pile-up
due to 
a high density. 
Dust particles composed by ballistic particle-cluster aggregation have 
$D \simeq 3$ but low effective densities relative to compact ones. 
Therefore, such porous particles with $D \simeq 3$ may produce high
enhancement factors due to large $x$ resulting from their low densities.  
In addition, high $x$ around a luminous star brings the
enhancement factors to increase with stellar luminosity, which is shown 
for dirty ice,
obsidian, and carbon in \citet{kobayashi08,kobayashi09}.

\begin{figure}[t]
\centerline{\includegraphics[width=7cm]{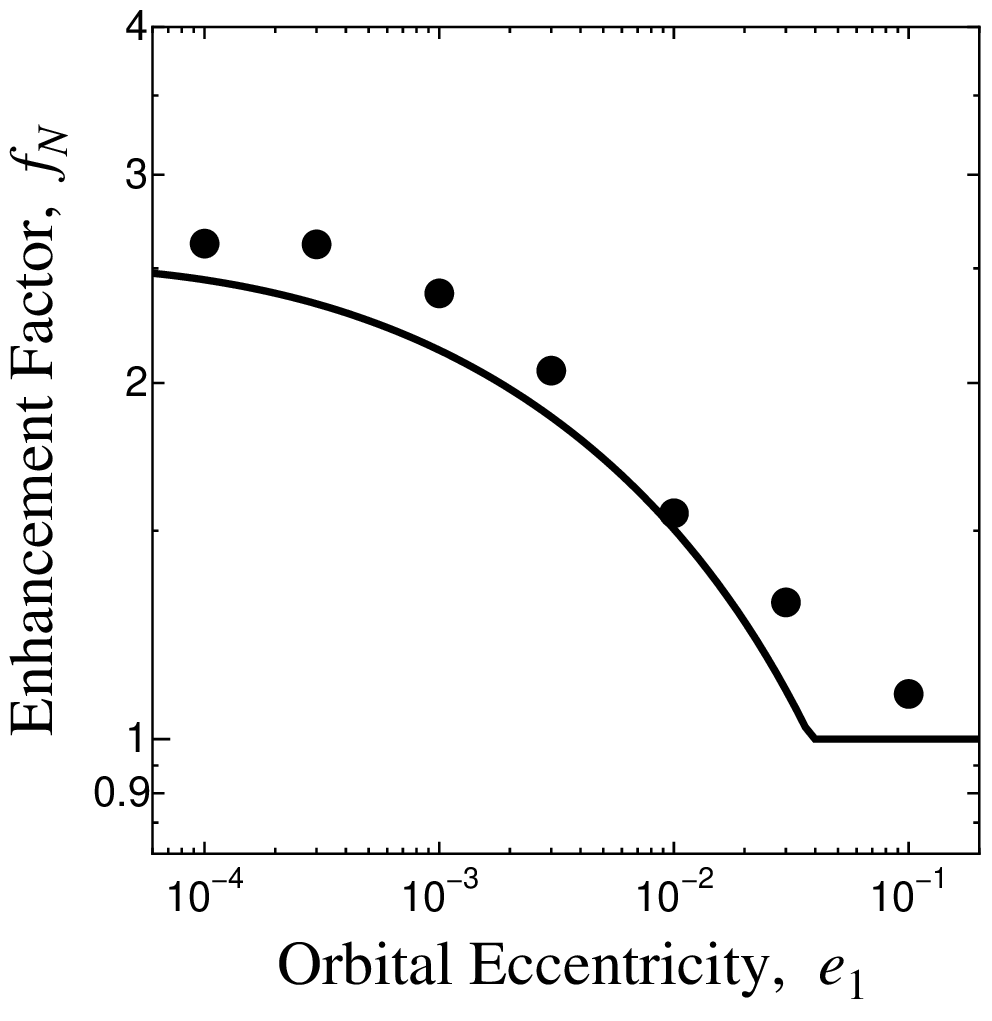}}
\caption{ 
Dependence of enhancement factor on $e_1$ for dirty ice, 
where $e_1$ is orbital eccentricities of dust particles at the beginning of
 their active sublimation. 
Solid line indicates Eq.~(\ref{eq:f_n}). 
Filled circles represent the results for the simulations calculated by
 \citet{kobayashi08}. 
\label{fig:dep_e1} }
\end{figure}

In \citet{kobayashi08}, we show the eccentricity dependence of enhancement
factors from our simulations.  The dependence is explained by $h(e_1)$
in the simplified formulae (see Fig.~\ref{fig:dep_e1}).  Dust particles
can pile up sufficiently for $e_1 \lesssim 10^{-3}$ because of $h(e_1) \simeq
1$.  Otherwise, the enhancement factors decrease with $e_1$.  For $e_1
\gtrsim 0.05$, the sublimation ring is not expected.

\section{Recipe}

We briefly show a recipe to obtain the sublimation temperature $T_{\rm
sub}$, its distance $a_{\rm sub}$, and the enhancement factors $f_N,
f_\tau$.  At first, the sublimation temperature $T_{\rm sub}$ is
available from Eq.~(\ref{eq:temp_app}) adopting the material properties
$\mu$, $H$, and $P_0$ listed in Table 1.  Then, we calculate the dimensionless
parameter $x$ through Eq.~(\ref{eq:x}), applying the stellar luminosity
and mass of interest and the bulk density listed in Table 1 for compact
spherical dust. Note that we should adopt a lower density for porous
particles, taking into account their porosity. 
 Inserting $x$ in Eq.~(\ref{eq:asub_app}), we derive the
sublimation distance $a_{\rm sub}$.  We further need orbital
eccentricities $e_1$ of dust particles at the beginning of active
sublimation to calculate the enhancement factors, $f_N$ and $f_\tau$.
The dust particles resulting from collisions have eccentricities $e_0
\sim \beta$ at the distance $a_0$ of the dust production region. Since
particles with the highest $\beta$ contribute most to a sublimation
ring, we estimate $e_0 \sim 0.5$.  Because eccentricities are dumped by
the P-R drag, we can calculate $e_1$ from the relation $a_{\rm sub}
e_1^{-4/5}(1-e_1^2) = a_0 e_0^{-4/5} (1-e_0^2)$ \citep{wyatt}.  Inserting $x$ and $e_1$
to Eqs. (\ref{eq:f_n}) and (\ref{eq:f_tau}), we obtain $f_N$ and
$f_\tau$.

\section{Discussion}

The asteroid belt and the Edgeworth-Kuiper belt (EKB) are possible dust sources in the solar
system. A dust counter on board spacecraft can measure the number density of dust 
particles.  The sublimation of icy dust occurs at $a_{\rm sub} =20$\,AU
given by Eq.~(\ref{eq:asub_app}). 
Icy particles reaching the sublimation zone from the EKB still have high
eccentricities $e_1\gtrsim 0.1$ \citep{kobayashi08}. 
Therefore, a
substantial sublimation ring is unexpected because of $f_N = 1$ for
$h(e_1) = 0$.  Since the number density of dust particles decreases
inside the sublimation zone, only a bump in the radial profile of the number
density appears around $a_{\rm sub}$ 
\citep[see ][]{kobayashi10}. 
In
contrast, dust
particles composed of rocky, refractory materials actively sublime at
several solar radii from the sun.  Therefore, orbital eccentricities of dust
particles coming from the asteroid belt drop to $\sim 0.01$ around the
sublimation distance due to the P-R drag.  Since we have $h(0.01) \simeq
0.3$ from Eq.~(\ref{eq:he1}) for $b = 11/6$ and $D=3$, $f_{N} \simeq
1.3$--3.0 is obtained form Eq.~(\ref{eq:f_n}) for $x \gtrsim 1$. Therefore,
future in-situ measurements of dust could find such a sublimation ring
of refractory dust particles originating from the asteroid belt, but not a
ring of icy dust particles from the EKB.

A dust ring was observed around $4 R_\sun$ from the sun in the period
1966-1983, although it was not detected in the 1990s 
\citep[][for a review]{kimura98}. 
The optical depth is measured by dust emission observations. 
The enhancement factor $f_{\rm \tau}$ of the observed ring is estimated to be
2--3 \citep{macqueen,mizutani}.  The mass distribution of drifting dust
particles for $b = 11/6$ is consistent with the measurement of dust
particles with masses ranging from $m_{\rm 0min} \sim 10^{-12}\,$g to $m_{\rm
0max} \sim 10^{-6}\,$g by spacecraft around the earth orbit 
\citep{grun}. 
Using
that, we estimate $f_{\tau} \lesssim 1.1$ from Eq.~(\ref{eq:f_tau}) for
$e_1 = 0.01$. 
Thus, the enhancement by sublimation cannot account for the observed dust
ring. However, this low $f_\tau$ is mainly caused by the mass range of
drifting dust particles. 
If $m_{\rm 0max}$ decreased during the transport of dust particles from 
the earth's orbit to the sublimation zone, higher $f_\tau$ could be
expected. For example, the largest particles in the mass distribution
become smaller by the sputtering from the solar wind. 
The sputtering may decrease $e_1$ as well as $m_{\rm 0max}$. 
Small particles with high eccentricities from the dust source are ground
down by sputtering and blown out by the radiation pressure before reaching the sublimation
zone, while large particles with low eccentricities gradually become small by sputtering without the increase of their eccentricities and drift into
the sublimation zone. 
If the ratio of $t_{\rm PR}$ to the timescale of decreasing size
due to sputtering ranges in 0.1--0.7, sublimation could form such a bright
ring because of small $m_{\rm 0max}$ and $e_1$. Indeed, the ratio
derived by 
\citet{mukai81} 
is 
consistent with the condition for the formation of a sublimation ring. 
That may be a clue to explain the observed ring. 

Debris disks found around main sequence stars would be formed through
collisional fragmentation in narrow planetesimal belts, which may
resemble the asteroids and Edgeworth-Kuiper belts in the solar system.  In young
debris disks, fragments produced by successive collisions are removed
from the disk by radiation pressure. 
We call such a disk a 
collision-dominated disk.  Once the amount of bodies has significantly
 been decreased through this process, the P-R drag becomes the main removal
process of fragments.  Such a disk is referred to as a drag-dominated
disk.  We have investigated the dust ring formation in drag-dominated
disks.
To observe a sublimation ring requires a high enhancement factor $f_\tau$
for the optical depth.  As shown in Eq.~(\ref{eq:f_tau}), a small
ratio of $m_{\rm 0max}$ to $m_{\rm 0min}$ yields high $f_\tau$. The
condition of $m_{\rm 0max} \sim m_{\rm 0min}$ is expected to form a
bright ring.  Since the drift time due to the P-R drag is comparable to
the collisional time for bodies with $m_{\rm 0max}$, the condition of
$m_{\rm 0max} \sim m_{\rm 0min}$ is achieved in transition from
a collision-dominated disk to a drag-dominated one. 

A significant sublimation ring consisting of icy particles is not
expected in a debris disk due to high eccentricities if a planetesimal belt as a dust
source is located within a few hundreds AU, similar to the solar system.  On the
contrary, dust particles composed of refractory materials have $e_1 \sim
0.01$ or smaller if a planetesimal belt is around the distance of the
asteroid belt or further outside.  Furthermore, refractory dust
particles have high
sublimation temperatures and hence produce a higher enhancement factor.
Recently, inner debris disks of refractory grains have been observed through interferometry
around Vega, $\tau$ Cet, $\zeta$ Aql, and $\beta$ Leo, and Formalhaut
\citep{absil06,absil08,difolco,akeson}. 
Such inner debris disks may have notable sublimation rings. 

To check the observability of a sublimation ring in the spectral energy distribution (SED) of
thermal emission expected from a disk around Vega located at the distance of
7.6\,pc from the earth, we take our formulae with $M_\star = 2.1
\,M_\sun$ and $L_\star = 59 \, L_\sun$ for compact spherical olivine particles. We obtain $T_{\rm sub} =
1300\,$K, $a_{\rm sub} = 0.35$\,AU and $f_{\rm \tau} = 3.1$, where we
adopt $m_{\rm 0max} = m_{\rm 0min}$ and $h(e_1) = 1$ in
Eq.~(\ref{eq:f_tau}).  The smallest radius 
$s_{\rm 0min} = 10\,\micron$ is much larger than the peak wavelength of
thermal emission ($\lambda_{\rm sub} \simeq 2.3\,\mu$m) and hence we simply treat dust particles as blackbodies 
to calculate the SED.  
The optical depth $\tau_{\rm d}$ of dust particles drifting by the P-R
drag without sublimation from the outer edge $a_{\rm
out}$ to the inner one $a_{\rm in}$ is given by a constant
$\tau_0$, where $a_{\rm in}$ equals $a_{\rm sub}$. 
The total optical depth $\tau$ of the
disk can be set as 
\begin{equation}
 \begin{array}{ccl}
 \tau &=& \tau_{\rm d} + \tau_{\rm e},\\
\tau_{\rm d} &=& \left\{
\begin{array}{cl}
\tau_0 & 
\quad \quad \quad \quad 
{\rm for} \quad a_{\rm in} < r < a_{\rm out}, \\
0 & 
\quad \quad \quad \quad 
{\rm otherwise, }
\end{array}
\right.\\
\tau_{\rm e} &=& \left\{
\begin{array}{lc}
\tau_0 (f_\tau -1)(a_{\rm sub}+\delta a_{\rm sub}-r)/\delta a_{\rm sub} 
& \\
\begin{array}{cl}
& \quad \quad \quad \quad 
 {\rm for} \quad a_{\rm sub} < r < a_{\rm sub}+\delta a_{\rm sub}, \\
0 & 
\quad \quad \quad \quad 
{\rm otherwise,}
\end{array}
\end{array}
\right. 
 \end{array}
\end{equation}
where $\tau_{\rm e}$ denotes 
the optical depth of dust particles
forming a dust ring with width $\delta a_{\rm aub}$.  
The
width of the ring is roughly given by $\delta a_{\rm sub} = 0.05
\,a_{\rm sub}$ \citep{kobayashi08,kobayashi09}.  We set $a_{\rm out} =
1.5 \, a_{\rm sub}$. Note that the other choice of $a_{\rm out}$ does not
change our result drastically. Choosing $\tau_0 = 8\times 10^{-3}$, we can reproduce the
flux density $\simeq 8.7\,$Jy measured by \citet{absil06} with the
interferometry at a wavelength of $2.1\,\micron$.\footnote[3]{\footnotemark[3]Note that $\tau_0$ depends on
the collision and drift timescales, $t_{\rm c}$, $t_{\rm PR}$, in a
planetesimal disk.  Drag-dominated disks should satisfy the condition
$t_{\rm c} \gtrsim t_{\rm PR}$; $\tau_0 \lesssim (1+\gamma) v_{\rm k}/c$ with the
Keplarian velocity $v_{\rm k}$ and the ratio $\gamma$ of the P-R drag
force due to the stellar wind to that due to the stellar
radiation.
The value of $\tau_0$ applied for fitting is much larger than that
for the drag-dominated disk ($\tau_0 \lesssim 3 \times 10^{-4}$), if we only
consider the P-R drag by the stellar radiation.  However, the mass loss
rate of Vega is estimated to be less than $3.4 \times 10^{-10} M_\sun
{\rm yr}^{-1}$ from radio-continuum observations 
\citep{hollis}. 
For the upper limit of the mass loss rate, drag-dominated disks can have the optical depth
$\tau_0 \lesssim 7 \times 10^{-2}$ because of the P-R drag due to the stellar
wind ($\gamma \sim 300$).  Thus the disk around Vega may be a drag-dominated disk.  
}   
Figure \ref{fig:sed} depicts that the sublimation ring does not
bring about a noticeable spectral feature in the SED, while the SED
strongly depends on $a_{\rm sub}$ value. 
The
flux density from the disk diminishes with decreasing wavelength for the
wavelength smaller than $\lambda_{\rm sub}$ 
because of the absence of dust particles with temperature higher than
$T_{\rm sub}$ due to sublimation.
The result for $a_{\rm in} = 0.6 a_{\rm sub} \simeq 0.21\,$AU is
shown to better 
agree with the observational data. Since $a_{\rm sub} \simeq 0.22$\,AU
($T_{\rm sub} \simeq 1700$\,K) for pyroxene, the Vega disk seems to be
abundant in pyroxene compared to olivine. 
That could be recognized as an evidence
of sublimation unless the light scattering of dust particles exceed
their thermal emission.\footnote[4]{\footnotemark[4]The scattering of light from  
the disk around Vega is negligible around the
wavelength $\sim 1\,\micron$ \citep{absil06}.}

\begin{figure}[t]
\centerline{\includegraphics[width=7cm]{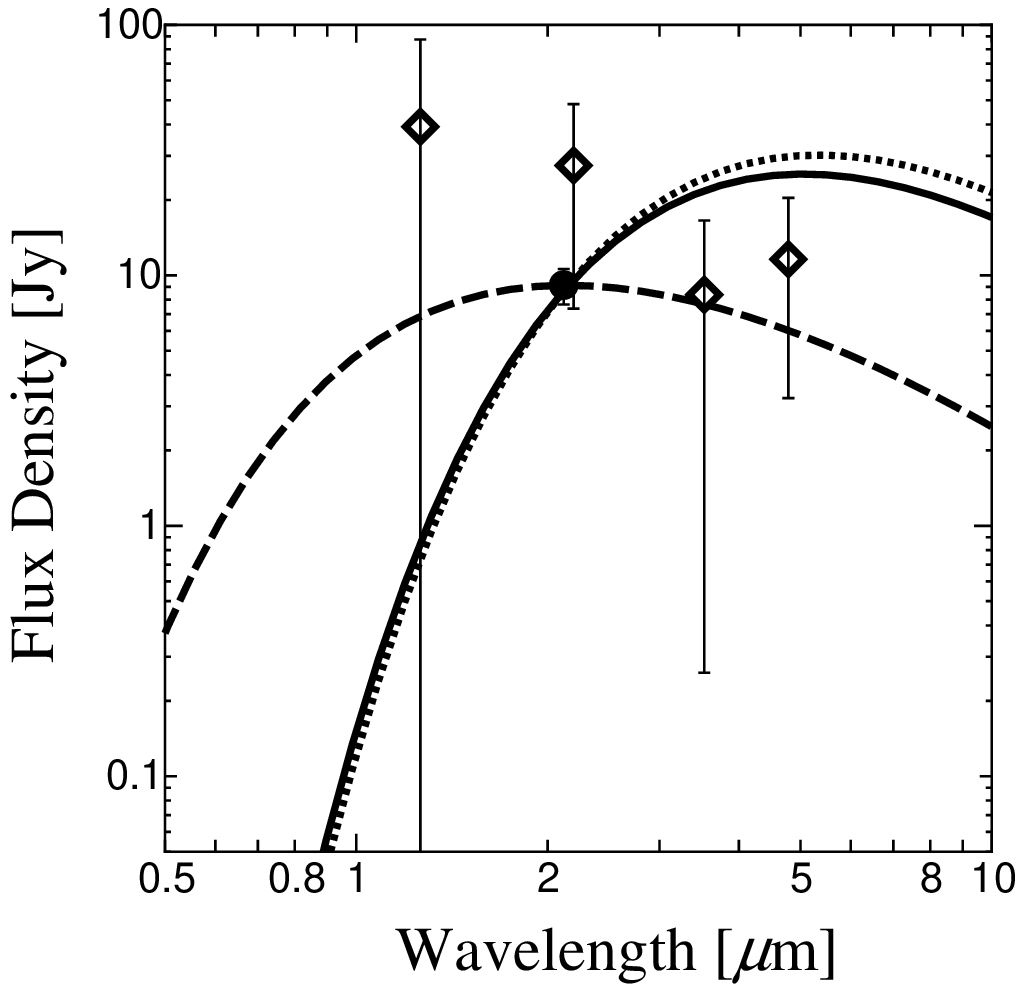}}
\caption{
The spectral density distribution of the disk around Vega. The solid line
 indicates the flux density from a disk with the sublimation ring,
 choosing $\tau_0 = 8\times 10^{-3}$ to fit the observational data at 2.1\,$\mu$m. Dotted lines
 represent that without a sublimation ring ($f_{\tau} = 1$ resulting in
 $\tau_{\rm e} = 0$) for $\tau_0 =1\times 10^{-2}$. 
To check the $a_{\rm sub}$ dependence, we set $a_{\rm in} = 0.6 a_{\rm sub}$ for $\tau_0
 =4\times 10^{-4}$ in the case without the
 sublimation ring (dashed lines). 
Circle corresponds to the interferometric measurement by
 \citet{absil06} and diamonds indicate the photometric data \citep[][and
 reference therein]{absil06}. \label{fig:sed} }
\label{fig:sed}
\end{figure}

\section{Summary}

\begin{itemize}
 \item[1.] We provide formulae of enhancement factors for the number
	   density and the optical depth due to a pile-up caused by sublimation
	   and its location and sublimation temperature, applying a 
	   simple model for optical cross sections of fractal dust particles. 
 \item[2.] High sublimation temperatures result in substantial enhancement
	   factors, though the pile-up is insensitive to the optical 
	   properties of dust particles. 
 \item[3.] If we adopt the mass distribution measured around the earth, 
	   the enhancement factor for the optical depth near the
	   sun is smaller than
	   1.1. Therefore, the enhancement cannot explain 
	   the solar dust ring detected in the epoch of
	   1966--1983, unless the 
	   largest particles were destroyed by sputtering.
 \item[4.] The number-density enhancement factor is expected to be
	   1.4--3 in the vicinity of the sun. 
	   Therefore, a sublimation ring could be found by in-situ
	   measurements by spacecraft around several solar radii 
	   from the sun.  
 \item[5.] Sublimation removes dust particles with temperatures higher	   
	   than their sublimation temperature. 
	   In the spectral energy distribution,  
	   the flux density from a disk reduces with decreasing wavelength, 
	   if the wavelength is shorter than the peak
	   one of the blackbody spectrum with the sublimation
	   temperature. 
	   That could be seen as a sublimation evidence. 
	   However, it is difficult to find signs from  
	   a sublimation ring in the spectral energy distribution. 
\end{itemize}

\appendix{Application limit}
\label{sc:limit} 

The value of $\beta$ increases with decreasing radius
as long as the radius fulfills the condition $s \gtrsim \lambda_\star$.  For
$s \lesssim \lambda_\star$, however, it decreases with decreasing
radius.\footnote[6]{\footnotemark[6]Note that 
$\beta$ is independent of $s$ for much smaller particles \citep{gustafson}.}
Hence, $\beta$ has a maximum value at $s \sim \lambda_\star =
(2898\,{\rm K}/T_\star) \, \micron$.  From Eq.~(\ref{eq:beta}), the
maximum value of $\beta$ is approximately given by
\begin{eqnarray}
 \beta_{\rm max} 
  &\sim& 0.98 \, \left(\frac{L_\star}{L_\sun} \right)\,
  \left(\frac{M_\sun}{M_\star} \right) 
\nonumber
\\
&& \times 
\left(\frac{T_\star}{5 \times 10^3\, {\rm K}}\right)
  \left(\frac{\rho}{1.0 \, {\rm g \,cm}^{-3}}\right)^{-1},      
\label{eq:est_beta} 
\end{eqnarray}
where the radiation pressure efficiency averaged over the stellar
radiation spectrum $\bar C_{\rm pr} \sim \pi s^2$ for $s \gtrsim \lambda_\star$. 
Here we define the effective density by 
\begin{equation}
 \rho = \frac{3m}{4\pi s^3}
\end{equation}
with the use of the dust mass $m$ and the characteristic radius $s$. 
Note that $\rho$ depends on $s$ in general; $\rho$ is constant for $m
\propto s^3$ (e.g., a compact sphere), whereas
$\rho$ is proportional to $s^{D\mbox{--}3}$ for a fractal aggregate with
fractal dimension $D$ 
\citep[see][ for the relation]{mukai92}. 

When dust particles are produced by successive collisions between large
bodies, their largest $\beta$ does not exceed $1/2$ because the dust with
$\beta > 1/2$ cannot resist against strong radiation pressure. 
These particles then drift into the sublimation zone. 
The smallest drifting dust particles that have the largest 
$\beta$ contribute most to the pile-up caused by sublimation. Because the
enhancement factors of the number density and the optical depth are
proportional to the largest $\beta$ value that 
drifting dust particles attain, we may expect insufficient pile-ups of
the particles if $\beta_{\rm max} < 1/2$.  We thus derive simplified
formulae for characteristics of a dust ring from the assumption of
$\beta_{\rm max} > 1/2$. Because the variation of $L_\star$ is much
larger than that of $T_\star$ for main sequence stars, this assumption
is translated into the condition for stellar luminosity given by
\begin{equation}
\frac{L_\star}{L_\sun} \gtrsim 0.5 \, \left(\frac{M_\star}{M_\sun} \right)
  \left(\frac{T_\star}{5 \times 10^{3} {\rm K}}\right)^{-1}	  
  \left(\frac{\rho}{1.0 {\rm g \, cm}^{-3}}\right), 
\label{eq:aplication_limit}
\end{equation}
if $\bar C_{\rm pr} \simeq \pi s^2$ for $s \gtrsim
\lambda_\star$.

\appendix{Derivation of enhancement factors}

We assume that the mass differential number of drifting dust particles
is proportional to $m_0^{-b+\eta}$
for the drifting particles with mass $m_0$.  
In addition, we adopt $\beta(m) \propto m^{-\eta}$, and $S(m) \propto
m^{\zeta_S}$, where $S$ is the geometrical cross section of a dust
particle with mass $m$. 
In \citet{kobayashi09}, 
we derived 
the number-density enhancement factor
$f_{{N}}$ and the optical-depth enhancement factor
$f_{\tau}$ at the peak as 
\begin{eqnarray}
 f_{N} &=&
\eta(m_{0{\rm min}})\,
  g_{\rm m}(T_{\rm sub},m_{0{\rm min}}) \,\,
  h_1(y_1,y_2)
  + 1,    
\label{eq:ns_or}
\\ \displaystyle 
 f_{\tau} &=&  
\eta(m_{0{\rm min}})\,
g_{\rm m}(T_{\rm sub},m_{0{\rm min}}) \,\, 
h_2(y_1,y_2)
	    + 1,     
\label{eq:tau_or}
\end{eqnarray} 
where
\begin{eqnarray}
 h_1 (y_1,y_2) 
&=& \frac{\int_1^{y_1} \tilde{y}^{-b} d \tilde{y}}
{\int_1^{y_2} \tilde{y}^{-b+\eta} d \tilde{y}},\label{eq:h1} \\
h_2(y_1,y_2) 
&=& \frac{\int_1^{y_1} \tilde{y}^{-b} d \tilde{y}}
{\int_1^{y_2} \tilde{y}^{-b+\eta+\zeta_S} d \tilde{y}},   
\label{eq:h2} 
\end{eqnarray}
$y_1 = m_{\rm init,max}/m_{\rm 0min}$, and 
$y_2 = m_{0{\rm max}}/m_{\rm 0min}$. 
As we will describe below, dust particles with $m_0 < m_{\rm init,max}$ drifting into 
the sublimation zone can contribute to the enhancement factors. 

Here $g_{\rm m}$ is a function of the optical properties of dust particles
with $m=m_{\rm 0min}$ at $T_{\rm sub}$, namely, given by a function of $x$. 
The function $g_{\rm m}$ is defined as 
\begin{equation}
 g_{\rm m}(T,m)
 = \frac{\frac{ 2 }{ 4 + c_{T}}
   \left(\frac{d \ln P_{\rm v}}{d \ln T} - \frac{1}{2} \right)- 2 }
   {1 + \frac{\eta}{1-\beta} - \eta - \zeta_A
     - \frac{d \ln \eta}{d \ln m} 
     - \left( \frac{d \ln P_{\rm v}}{d \ln T} 
     - \frac{1}{2} \right)
       \frac{c_\star - c_{\rm d } 
       }
       {4 + c_{T}}}, 
\label{defgm(T.m)}
\end{equation}
where
\begin{eqnarray}
 c_\star = \frac{d \ln \bar C_\star}{d \ln m}, \quad
 c_{\rm d} = \left(\frac{\partial \ln \bar C_{\rm d}}
                        {\partial \ln m}\right)_T, 
\nonumber
\\
 c_{T} = \left(\frac{\partial \ln \bar C_{\rm d}}
                    {\partial \ln T}\right)_m, \quad
 \zeta_A = \frac{d \ln A}{d \ln m}.  
\end{eqnarray}
According to our simple model in 
Sections~\ref{sc:temp} and \ref{sc:fractal_dust}, 
a particle with $T_{\rm sub}$ and $m_{\rm 0min}$ has
\begin{eqnarray}
c_\star &=& 2/D, \quad c_{\rm d} =
2/D + 1/(x+1)D, \quad \zeta_A = \zeta_S = 2/D, 
\nonumber
\\
\eta &=& - d \ln \beta / d \ln m = (D-2)/D, \quad \beta = 1/2. 
\label{eq:fractal_dust_model}
\end{eqnarray}
Then, $g(T_{\rm sub},m_{\rm 0min})$ reduces to 
\begin{equation}
 g_{\rm m}(T_{\rm sub}, m_{\rm 0min}) = \frac{2 I D (1+x)}{2\alpha (1+x) + I },
\label{eq:gm_ap}
\end{equation}
where $c_T = 0$, and $4I = d \ln P_{\rm v} /d \ln T \gg 1$.  We define
$\eta g_{\rm m}(T_{\rm sub}, m_{\rm 0min})$ as $g(x)$ given by
Eq.~(\ref{eq:gfunc}).

Only the dust particles with initial masses ranging from $m_{\rm 0min}$
to $m_{\rm init, max}$ can stay long around the distance $a_{\rm sub}$
for a pile-up.  Large particles initially pass the distance $a_{\rm
sub}$ and approach there again by outward drift due to active
sublimation \citep[see Fig. 1 of ][]{kobayashi08}. 
Orbital
eccentricities $e$ of the particles rise during the active
sublimation. If $e \gtrsim 2 kT_{\rm sub}/\mu m_{\rm u}H$, they are blown
out immediately and hence do not contribute to the 
formation of a dust ring.
Therefore, $m_{\rm init, max}$ depends on the eccentricity $e_1$ of a
dust particle starting the active sublimation.  
\citet{kobayashi09} derive the relation between $e$ and $m$ during the active sublimation as
\begin{equation}
 e = \left(\frac{1-\beta(m_1)}
                {1-\beta(m) }\right)^{\kappa} e_1, 
\label{eq:zeroe}
\end{equation}
where 
\begin{equation}
 \kappa = \frac{1}{4 + c_T}
     \left(\frac{d \ln P_{\rm v}}{d \ln T} - \frac{1}{2}\right)
   - \frac{5}{4}
\label{eq:kappa}, 
\end{equation}
with $m_1$ is the mass starting the active sublimation.\footnote[5]{\footnotemark[5]Eq.~(\ref{eq:kappa}) is different from Eq.~(57)
in \citet{kobayashi09} 
because $d \ln T/ d \ln a$ in their Eq.~(29)
should be replaced by $\partial \ln T/\partial \ln a$.  Then, we obtain
Eq.~(\ref{eq:kappa}) instead of their Eq.~(57).}  We approximate $\kappa
= I$ because $I \gg 1$.  Equation~(\ref{eq:zeroe}) indicates that $e$
substantially changes for $\beta \sim 1$.  Since the mass loss is
negligible outside the sublimation zone, we approximate $m_1\simeq m_0$.  Particles can pile up as long as $e \lesssim 2kT_{\rm sub}/\mu
m_{\rm u} H = 1/2I$ \citep{kobayashi09}. 
Substituting $e = 1/2I$,
$\beta = 1/2$, and $\beta_1 = \beta (m_{\rm init,max}/m_{\rm
0min})^{-\eta}$ into Eq.~(\ref{eq:zeroe}), we have
\begin{equation}
 m_{\rm init,max} = \left[\frac{(2 I e_1 )^{1/I}}{2(2 I e_1 )^{1/I}-1} \right]^{1/\eta}
m_{\rm 0min}. \label{eq:rp} 
\end{equation}
Equation~(\ref{eq:rp}) is valid for $m_{\rm init,max} \leq m_{\rm 0max}$
and $ e_1 \geq 1/ I 2^{I+1}$.  We should set $m_{\rm init,max} = m_{\rm
0max}$ instead of Eq.~(\ref{eq:rp}) for $m_{\rm init,max} > m_{\rm
0max}$ or $ e_1 < 1/ I 2^{I+1}$.  Because $y_2 \gg 1$, and $b =
1.8$--1.9 in Eqs (\ref{eq:h1}) and (\ref{eq:h2}), $h_1$ and $h_2$ are
given by
\begin{eqnarray}
h_1(y_1,y_2) &=& 
\frac{b-\eta-1}{b-1}\,\,
h(e_1), 
\label{eq:h1_new} 
\\
h_2(y_1,y_2) &=& \frac{2-b}{b-1} 
\left(\frac{m_{\rm 0min}}{m_{\rm 0max}}\right)^{2-b}
h(e_1), 
\label{eq:h2_new} 
\end{eqnarray}
where $h$ is defined as Eq.~(\ref{eq:he1}). 

Substituting Eqs.~(\ref{eq:gm_ap}),  (\ref{eq:h1_new}), and (\ref{eq:h2_new}) 
into Eqs.~(\ref{eq:ns_or}) and
(\ref{eq:tau_or}), 
we have the enhancement factors in Eqs.~(\ref{eq:f_n}) and
(\ref{eq:f_tau}). 

\acknowledgments{We appreciate the advice and encouragement of
A. Krivov, M. Ilgner, and M. Reidemeister. The careful reading of the
manuscript by the anonymous reviewers helps its improvement. This research is supported by grants from CPS, JSPS, and MEXT Japan.}

\email{Hiroshi Kobayashi (e-mail: hkobayas@astro.uni-jena.de)}
\label{finalpage}
\lastpagesettings
\end{document}